\documentclass[fleqn,10pt]{wlscirep}
\usepackage{hyperref}
\usepackage{float}
\usepackage[super]{nth}
\usepackage{multirow}
\usepackage{soul}
\usepackage{cite}
\usepackage{xcolor}
\usepackage{algorithm}
\usepackage{algpseudocode}
\usepackage{makecell}
\usepackage{booktabs}
\usepackage{titlesec}

\title{Tracing Cross-chain Transactions between EVM-based Blockchains: An Analysis of Ethereum-Polygon Bridges}

\author[1,2*]{Tao Yan}
\author[3]{Chuanshan Huang}
\author[1,2]{Claudio J. Tessone}

\affil[1]{Blockchain \& Distributed Ledger Technologies, Department of Informatics, University of Zurich, Z\"urich, Switzerland}
\affil[2]{UZH Blockchain Center, University of Zurich, Z\"urich, Switzerland}
\affil[3]{Advanced Blockchain AG, Germany}
\affil[*]{yan@ifi.uzh.ch}


\begin{abstract}
Ethereum’s scalability has been a major concern due to its limited transaction throughput and high fees. To address these limitations, Polygon has emerged as a sidechain solution that facilitates asset transfers between Ethereum and Polygon, thereby improving scalability and reducing costs. However, current cross-chain transactions, particularly those between Ethereum and Polygon, lack transparency and traceability. This paper proposes a method to track cross-chain transactions across EVM-compatible blockchains. It leverages the unique feature that user addresses are consistent across EVM-compatible blockchains. We develop a matching heuristic algorithm that links transactions between the source and target chains by combining transaction time, value, and token identification. Applying our methodology to over 2 million cross-chain transactions (August 2020–August 2023) between Ethereum and Polygon, we achieve matching rates of up to 99.65\% for deposits and 92.78\% for withdrawals, across different asset types including Ether, ERC-20 tokens, and NFTs. In addition, we provide a comprehensive analysis of various properties and characteristics of cross-chain transactions. Our methodology and findings contribute to a better understanding of cross-chain transaction dynamics and bridge performance, with implications for improving bridge efficiency and security in cross-chain operations.
\end{abstract}
\begin{document}

\flushbottom
\maketitle
\thispagestyle{empty}
tion{Introduction}
Ethereum has facilitated the development of blockchain by enabling smart contracts. However, Ethereum’s scalability limitations\cite{bez2019scalability,dennis2019analysis,yang2020review,scherer2017performance} and high transaction fees\cite{de2021analysis,donmez2022transaction,zhao2023cost} have been major issues. Currently, Ethereum can only process around 12-15 transactions per second on average \cite{Whatdoes30:online}. To address these issues, Layer 2 scaling solutions have emerged as additional mechanisms utilized outside of the blockchain system to improve Ethereum’s performance. \cite{sguanci2021layer,hafid2020scaling}. Among them, Polygon (formerly known as Matic)\cite{kanani2021matic} stands out as one of the most prominent solutions. Polygon enhances Ethereum’s efficiency by serving as a sidechain\cite{singh2020sidechain} where assets can be transferred between blockchains at lower costs and higher speeds.

Despite the growing adoption of Layer 2 scaling solutions, tracking and analysing cross-chain transactions remains a significant challenge. Currently, only users who initiate transactions can access information regarding cross-chain transactions, limiting transparency for the broader public. This lack of transparency hinders the tracking of asset flows across blockchains, posing challenges for auditing, compliance, and risk management—especially in contexts requiring AML (anti-money laundering) oversight\cite{kemal2014anti}. Moreover, a deeper understanding of asset flow patterns, user behaviour, and bridge performance is essential to improve transparency and the design of more robust cross-chain protocols.  These limitations highlight the need for a general and reliable method to track and analyse cross-chain transactions.

This work proposes a heuristic algorithm for tracing cross-chain transactions between EVM-compatible blockchains. We systematically investigate the cross-chain mechanisms and propose a novel method to match these transactions. We use the Ether-Polygon bridge as an example to apply and evaluate the algorithm’s performance in a real-world context. Additionally, we provide an in-depth empirical analysis of cross-chain activity between Ethereum and Polygon. Our work offers the following contributions:
\begin{enumerate}
    \item We propose a matching heuristic algorithm that can trace cross-chain transactions between EVM-compatible blockchains, achieving high accuracy rates across different asset types. The methodology not only improves asset traceability but is also extendable to other EVM-compatible blockchain pairs.
    \item We uncover significant temporal asymmetries in cross-chain operations, with transfers from Ethereum to Polygon completing substantially faster than those in the reverse direction. Furthermore, we find that the Ethereum Merge notably prolonged cross-chain deposit completion times.
    \item We conduct a large-scale analysis of over 2 million matched cross-chain transactions, yielding several key insights: (1) Asset flows are predominantly one-way from Ethereum to Polygon, with withdrawal rates generally below 50\%. However, this trend temporarily reversed around the time of the Ethereum Merge. (2) Cross-chain ERC-20 activity is dominated by stablecoins. (3) NFT transaction activity exhibits counter-intuitive patterns where Polygon’s fee advantage has yet to drive substantial activity away from Ethereum.
    \item We identify three potential security risks: (1) unlike deposits, which require only one transaction, withdrawals involve two steps. This design inconsistency can cause users to overlook the second action, leaving assets unclaimed and potentially leading to loss; (2) centralisation in bridge governance, as the Polygon PoS bridge relies on a relatively small validator set (approximately 105) with the authority to control bridge operations; and (3) prolonged cross-chain settlement times, with some transactions taking up to 6 months to complete, exposing the system to attack vectors during these extended confirmation windows.
\end{enumerate}

The paper is organized as follows: In Section \ref{sec:background}, we present the background information on Ethereum, Polygon, and cross-chain bridges, specifically focusing on those operating on Polygon. Section \ref{sec:related_work} presents the related works on cross-chain research. Section \ref{sec:method} outlines our proposed method, elucidating the fundamental concept behind matching transactions and the algorithm used for transaction matching. Section \ref{sec:data} discusses the data collection process, while Section \ref{sec:results} presents the comprehensive results obtained. Section \ref{sec:dis} discusses extensions of proposed algorithm and the bridge governance structures. Finally, Section \ref{sec:con} concludes this work and discusses the limitations and future research directions.

\section{Background}\label{sec:background}
This section outlines key concepts essential for understanding the cross-chain mechanism, with a specific focus on the Polygon PoS Bridge mechanism.

\subsection{Ethereum}
Ethereum is the first blockchain that introduced smart contracts\cite{buterin2014next}. Its purpose is to exceed the conventional capabilities of Bitcoin by enabling developers to build decentralized applications (DApps) and execute programmable transactions. It introduced the Ethereum Virtual Machine (EVM)\cite{hirai2017defining}, a virtual environment that provides the runtime infrastructure for executing smart contracts. Although Ethereum has greatly facilitated the development of blockchain ecosystems, its scalability has raised significant concerns due to its limited transaction throughput and elevated transaction fees.
\subsection{Polygon}
Polygon is a Layer 2 scaling solution designed to improve Ethereum’s scalability and efficiency. It functions as a sidechain to Ethereum, facilitating faster transaction processing with lower fees. Processed transactions are periodically submitted to the Ethereum mainnet for finality and verification~\cite{PolygonA52:online}. Polygon supports various Layer 2 technologies, including Plasma Chains, zkRollups, and Optimistic Rollups, each employing different strategies to enhance scalability. By leveraging these solutions, users can benefit from faster transactions and lower fees while maintaining the ability to interact with the Ethereum ecosystem.
\subsection{Cross-chain mechanism between Ethereum and Polygon}
\begin{figure}[htbp]
\centerline{\includegraphics[scale=0.5]{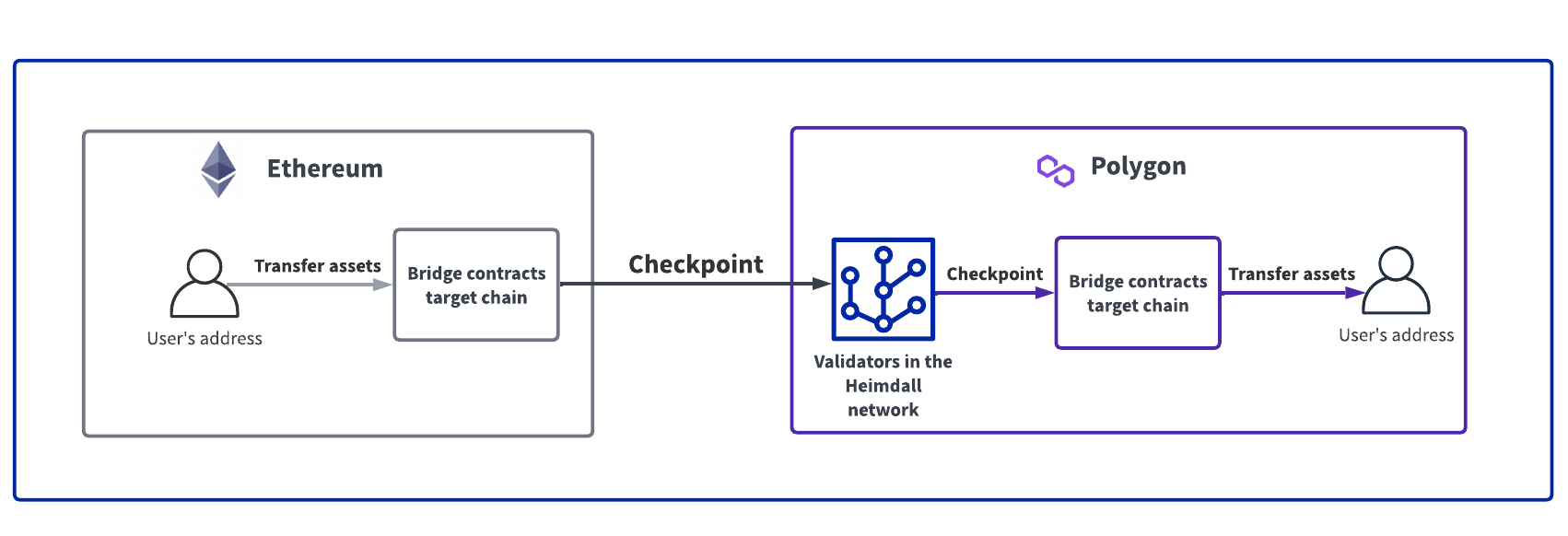}}
\caption{An overview of cross-chain transactions between Ethereum and Polygon.}
\label{crosschaingeneral}
\end{figure}

A cross-chain transaction is not a single transaction by nature, instead, it is composed of two transactions on each blockchain. When users initiate a cross-chain transaction, they first transfer their tokens to a bridge contract on the source chain through a transaction. Upon confirming the receipt of the user's tokens, the bridge contract then transmits the data to the target chain. Subsequently, the bridge contract on the target chain receives the data from the source chain and proceeds to initiate the second transaction to send the assets to the user's address on the target chain. Notably, the user address that transfers the asset on the source chain is the same as the user address that receives the asset on the target chain in EVM-based blockchain systems. Figure \ref{crosschaingeneral} illustrates this process.

\subsection{Mechanism of Polygon PoS Bridge}
\begin{figure}[htbp]
\centerline{\includegraphics[scale=0.75]{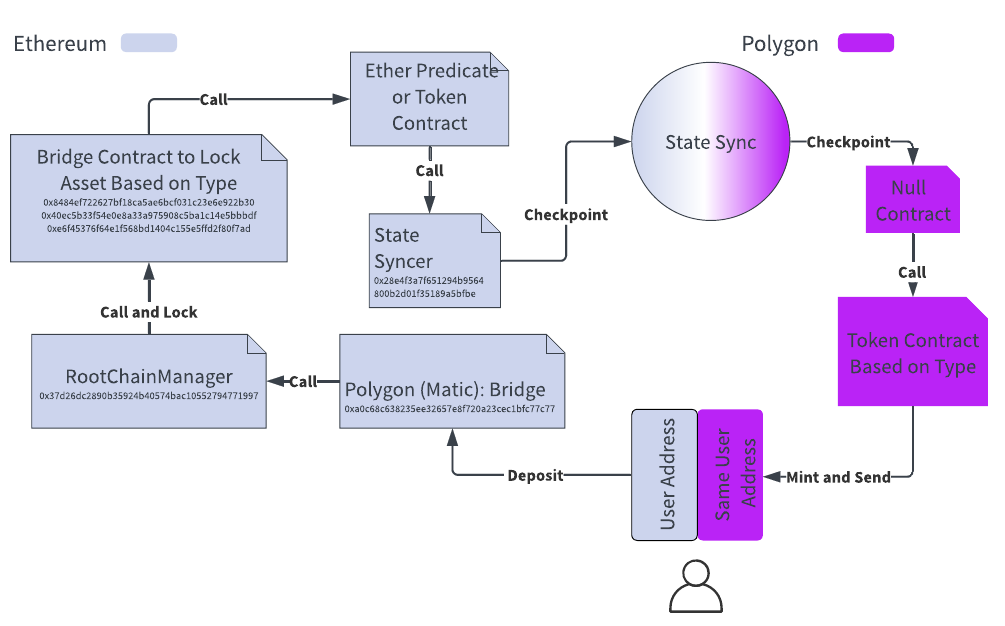}}
\caption{The Mechanism of Polygon PoS Bridge.}
\label{posmechanism}
\end{figure}

There are five official bridges between Ethereum and Polygon PoS, each of which is a smart contract deployed on Ethereum. Three of these contracts are responsible for locking Ethers, ERC20 tokens, or ERC721 tokens, which are summarized in Table \ref{bridgecontract}. These three contracts are invoked by the main bridge contract, \emph{Polygon (Matic): Bridge}, through a chain of contract calls as illustrated in Figure \ref{posmechanism}. Additionally, there is a \textit{Polygon (Matic): Plasma Bridge\footnote{0xA0c68C638235ee32657e8f720a23ceC1bFc77C77}} which can transfer all types of assets. However, due to its relatively low usage compared to the main bridge, this work does not focus on it.

\textbf{Lock and mint} As illustrated in Figure \ref{posmechanism}, when a user initiates a transaction to transfer assets from Ethereum to Polygon, they must first approve and send the assets to the main bridge contract which is \emph{Polygon (Matic): Bridge} on Ethereum. Then, this contract initiates a series of contract calls, to lock the user's assets. This process emits events such as \emph{LockedEther()}, 
\emph{LockedERC20()}, \emph{LockedERC721()} and \emph{LockedERC721Batch()} based on the type of assets being transferred. The last step on Ethereum is that a contract called \emph{State Syncer} will be called to transmit \emph{State Synchronization} data to Polygon. On Polygon's side, once the \emph{State Synchronization} data is updated in the data layer, Polygon's \emph{Null Contract} will call the respective token contract to mint the desired amount of assets. Finally, the minted assets will be sent to the user's address on Polygon.

\begin{table*}[htbp]
\caption{An overview of the bridge contract on Ethereum for the Polygon PoS Bridge.}
\centering
\begin{tabular}{l l l}
\toprule
\textbf{Bridge Contract on Ethereum} & \textbf{Asset Type} & \textbf{Event} \\
\midrule
Polygon (Matic): Ether Bridge & Ether & LockedEther() \\
Polygon (Matic): ERC20 Bridge & ERC20 tokens & LockedERC20() \\
\multirow{2}{*}{Polygon (Matic): ERC721 Predicate Proxy} & \multirow{2}{*}{ERC721 tokens} & LockedERC721Batch() \\
& & LockedERC721() \\
\bottomrule
\end{tabular}
\label{bridgecontract}
\end{table*}
\textbf{Burn and withdraw} When a user initiates a transaction that will transfer assets from Polygon to Ethereum, the process involves initially sending the assets to the \emph{Null Contract}, also referred to as burning the assets. Proof of this event must be submitted to the Ethereum bridge contract to unlock the original tokens. This process requires checkpointing. Polygon periodically sends block summaries (checkpoints) to Ethereum via a set of validators. These validators are responsible for validating and signing checkpoints. Subsequently, on Ethereum, the locked assets are returned to the users' addresses. More specifically, locked Ethers and ERC721 tokens are sent back to the user through transaction events known as  \emph{ExitedEther()} and \emph{ExitedERC721()} in Table \ref{withdrawalcontract}, respectively. However, locked ERC20 tokens are simply transferred by utilizing the default \emph{Transfer()} events of ERC-20 token contracts.
\begin{table*}[htbp]
\caption{An overview of the withdrawal process contracts and events on Ethereum.}
\centering
\begin{tabular}{l l l}
\toprule
\textbf{Contract on Ethereum} & \textbf{Asset Type} & \textbf{Event} \\
\midrule
Polygon (Matic): Ether Bridge  & Ether & ExitedEther() \\
Polygon (Matic): ERC721 Predicate Proxy & ERC721 tokens & ExitedERC721() \\
ERC20 Token Contracts  & ERC20 tokens & Transfer() \\ 
\bottomrule
\end{tabular}
\label{withdrawalcontract}
\end{table*}

\section{Related works}\label{sec:related_work}

Ou et al.\cite{ou2022overview} provide a comprehensive review of blockchain cross-chain technology, analyzing mainstream solutions, comparing their advantages and disadvantages, and identifying key challenges and potential solutions. Shadab et al.\cite{shadab2020cross} propose a uniform protocol for general cross-chain transactions and develop XCHAIN that can automatically generate cross-chain transactions. Chervinski et al. \cite{chervinski2023analyzing} focus on the performance evaluation of cross-chain communication, specifically the Cosmos Inter-Blockchain Communication (IBC) Protocol. Their empirical evaluation framework highlights major scalability bottlenecks, including high transaction latency and concurrency limitations. Cao et al. \cite{cao2023cross} propose a cross-chain traceability mechanism using notary groups and reputation-based elections to enhance trust and secure data exchange.

Layer 2 blockchain protocols have emerged as a key solution for improving blockchain scalability. Several Layer 2 solutions, including ZK Rollup, Validium, Optimistic Rollup, and Optimium, have been proposed to enhance efficiency. Song et al. \cite{mandal2023investigating} compare Layer 1 and Layer 2 scaling approaches, analyzing the storage and latency limitations of Layer 1 and exploring the benefits of Layer 2 solutions like rollups and payment channels. Singh \cite{singh2020sidechain} examines the role of sidechains in improving scalability, privacy, and security while also addressing implementation challenges and proposing potential improvements. While most of the cross-chain transaction research focuses on Layer 2 solutions, some studies explore Layer 0 interoperability. Zarick et al. \cite{zarick2021layerzero} introduce LayerZero, the first trustless omnichain interoperability protocol, which provides a low-level communication primitive to enable diverse cross-chain applications. They also propose the $\Delta$ (Delta) algorithm \cite{TheBridg5:online}, a novel resource-balancing methodology that utilizes cross-chain liquidity to facilitate native-asset-based cross-chain transactions with instant finality. Huang et al. \cite{huang2024seamlessly} analyze the architecture and operational dynamics of the Stargate Layer-0 cross-chain bridge, uncovering vulnerabilities and evidence of exploitations.

Current research predominantly focuses on the design and mechanisms of cross-chain solutions. However, empirical studies on real-world cross-chain transactions remain limited, highlighting a gap in understanding their practical performance and security.

\section{Methodologies}\label{sec:method}
This section presents our algorithm for identifying cross-chain transactions between EVM-compatible blockchains.
% \iffalse
\subsection{Address consistency in EVM-based blockhains} In EVM-based blockchains, the fundamental mechanism for tracing cross-chain transactions relies on using the same external address across different blockchains. The feature of "Same Address for All EVM Chains" ensures consistency in the addresses generated for Ethereum and other EVM-supported chains\cite{SameAddr56:online}. For instance, an external address created on Ethereum can be seamlessly used by the same user on other EVM-based blockchains, such as Polygon, and Avalanche\cite{rocket2019scalable}. When users initiate cross-chain transfers via bridges, their digital assets, including native cryptocurrencies and tokens, are moved to the same address on the target EVM-compatible chain. Therefore, the reconciliation of cross-chain transactions can be achieved by identifying a user's transaction on one blockchain and subsequently locating its corresponding transaction on another blockchain. This matching process involves locating transactions from the same address on both chains and aligning them based on factors such as timestamp, value, and token identification.

\subsection{Heuristic algorithm to match cross-chain transactions}\label{secmatch_algo}  

To identify the transactions that traverse across blockchains, we need to manually match the data since there is no explicit connection for transactions that travel between Ethereum and Polygon, even not on Polygon's bridge explorer
\footnote{https://bridge-explorer.polygon.technology/deposits}. To address this, we have developed a heuristic algorithm to match the cross-chain transactions. We formally define the algorithm as follows. Let \emph{E} represent the set of transactions on Ethereum, and \emph{P} represent the set of transactions on Polygon. Each data field within the sets will be denoted as DataSet.DataField. For example, timestamps in Ethereum data will be denoted as \emph{E.timestamp}. Data is matched based on their token type and four criteria as shown in Table \ref{matchingtable}. Criterion 1 ensures that the deposited assets are received by the same address; criterion 2 reinforces that the deposit shall be received within a reasonable time window, which is characterized by a tunable parameter \emph{time\_tolerance}. Criterion 3 ensures that token have the same names (or Ether for WETH), and Criterion 4 checks if the source transaction and target transaction are transmitting the same amount of assets. 

Algorithm~\ref{algorithm1} outlines the pseudocode of our heuristic matching procedure. For each Ethereum transaction, we identify candidate Polygon transactions based on the deposit receiver address, then retain only those satisfying all applicable criteria.
\\

\begin{algorithm}
\caption{Cross-chain Deposit Match Heuristics}\label{alg:cap}
\small
\begin{algorithmic}[1]
\For{each transaction \emph{e} in \emph{E}} 
    \State Get\_Data(\emph{p}) \emph{s.t.} \emph{p} in \emph{P} is the list of transactions for e.depositReceiver

\If{e.token = Ether}{
    \State tmp = join(e, p)
    \If{(p.token$\neq$ WETH) or (p.timestamp - e.timestamp $\geq$ time\_tolerance) or e.value$\neq$p.value }
        \State tmp.drop\_rows()
    \EndIf
}
    
\ElsIf{e.token\_type = ERC20}
    \State tmp = join(e, p)
    \If{(e.token\_name $\neq$ p.token\_name) or ( p.timestamp - e.timestamp $\geq$ time\_tolerance) or (e.value$\neq$p.value) }
        \State tmp.drop\_rows() 
    \EndIf
\Else{ e.token\_type = ERC721}
    \State tmp = join(e, p)
    \If{(e.token\_name $\neq$ p.token\_name) or (e.token\_id $\neq$ p.token\_id) or (p.timestamp - e.timestamp $\geq$ time\_tolerance)} 
        \State tmp.drop\_rows() 
    \EndIf

\EndIf
\EndFor
\State return tmp
\end{algorithmic}
\label{algorithm1}
\end{algorithm}

\begin{table*}[htbp]
\caption{Matching criteria based on asset type}
\centering
{\small
\begin{tabular}{c c c c c}
\toprule
\textbf{Token Type} & \textbf{Criteria 1} & \textbf{Criteria 2} & \textbf{Criteria 3} & \textbf{Criteria 4} \\
\midrule
Ether & 
\makecell{E.depositReceiver \\= P.to\_address} & 
\makecell{$|$E.timestamp - P.timestamp$|$\\$\leq$ time\_tolerance} & 
\makecell{P.token\_address \\ = `WETH'} & 
\makecell{E.value \\= P.value} \\
\midrule
ERC20 & 
\makecell{E.depositReceiver \\= P.to\_address} & 
\makecell{$|$E.timestamp - P.timestamp$|$\\$\leq$ time\_tolerance} & 
\makecell{P.token\_symbol \\= E.token\_symbol} & 
\makecell{E.value \\= P.value} \\
\midrule
ERC721 & 
\makecell{E.depositReceiver \\= P.to\_address} & 
\makecell{$|$E.timestamp - P.timestamp$|$\\$\leq$ time\_tolerance} & 
\makecell{P.token\_symbol \\= E.token\_symbol} & 
\makecell{E.token\_ID \\= P.token\_ID} \\
\bottomrule
\end{tabular}
}
\label{matchingtable}
\end{table*}

\begin{figure}[htbp]
\centerline{\includegraphics[scale=0.5]{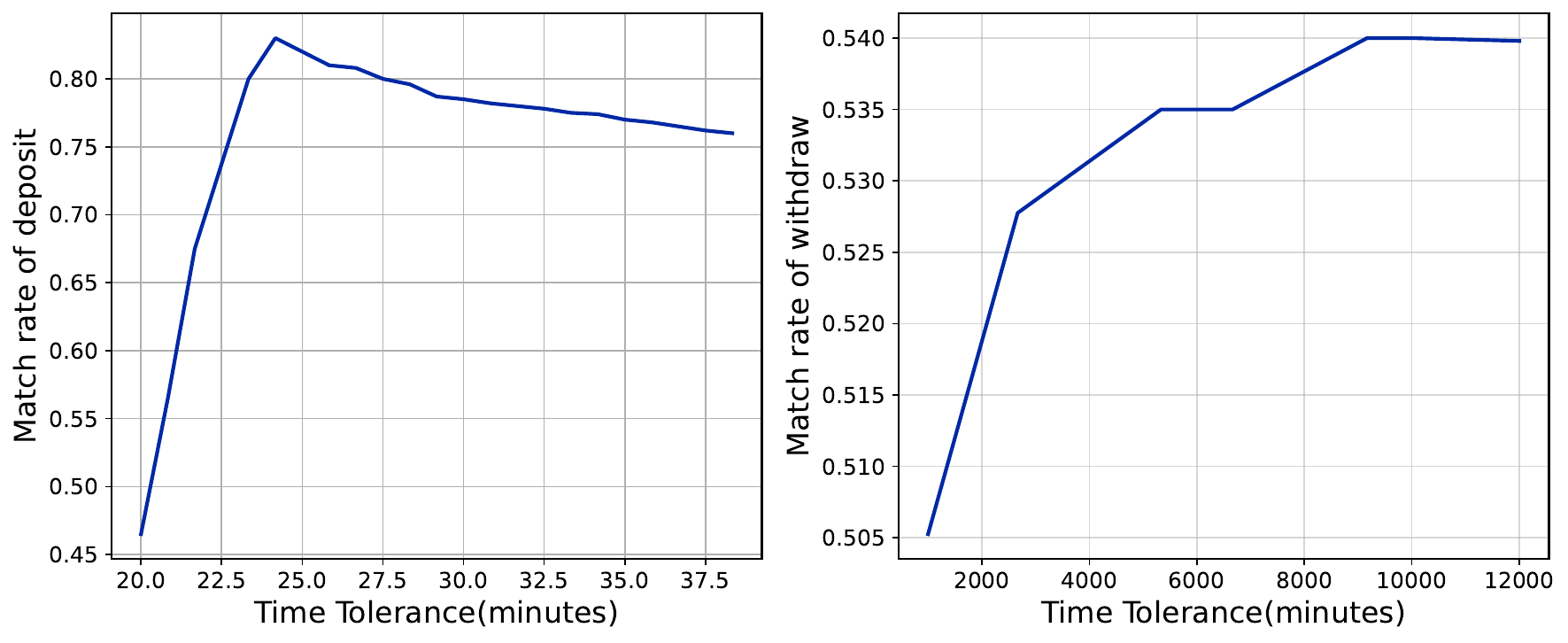}}
\caption{Match rate under different \emph{time\_tolerances}. The left subgraph represents the match rate for deposit transactions, while the right subgraph represents the match rate for withdrawal transactions. }
\label{matchrate}
\end{figure}

In practice, the value of the parameter \emph{time\_tolerance} needs to be manually determined. To determine the optimal value, we randomly sample 10,000 cross-chain transactions to test the match rate of our algorithm. We run the algorithm with different values of \emph{time\_tolerance} and record the percentage of an exact match. An exact match is defined as a match where a transaction from Ethereum is matched to only one transaction on Polygon. If a record does not match any Polygon record or matches with multiple Polygon records, it is not considered an exact match. Therefore, an exact match is a valid cross-chain transaction. As shown in Figure \ref{matchrate}, for deposits, the match rate increases sharply with larger \texttt{time\_tolerance} values, peaking at around 24.2 minutes before gradually declining since more duplicated yet incorrect matches are included with a higher time threshold. On the other hand, the withdrawal match rate continuously increases until it reaches its peak at around 9,166.7 minutes as shown in the right panel. It is worth noting that the \texttt{time\_tolerance} value was derived from a random sample of 10,000 transactions, and thus reflects an upper-bound estimate of transaction latency rather than the average time cost. Additionally, the match rates shown in Figure~\ref{matchrate} represent performance on the sampled dataset only and do not capture the match rates of the full transactions. Detailed matching rates by asset type are reported in Table~\ref{matchingratetable}.

\section{Data Collection}\label{sec:data}

\begin{figure}[h]
\centerline{\includegraphics[scale=0.7]{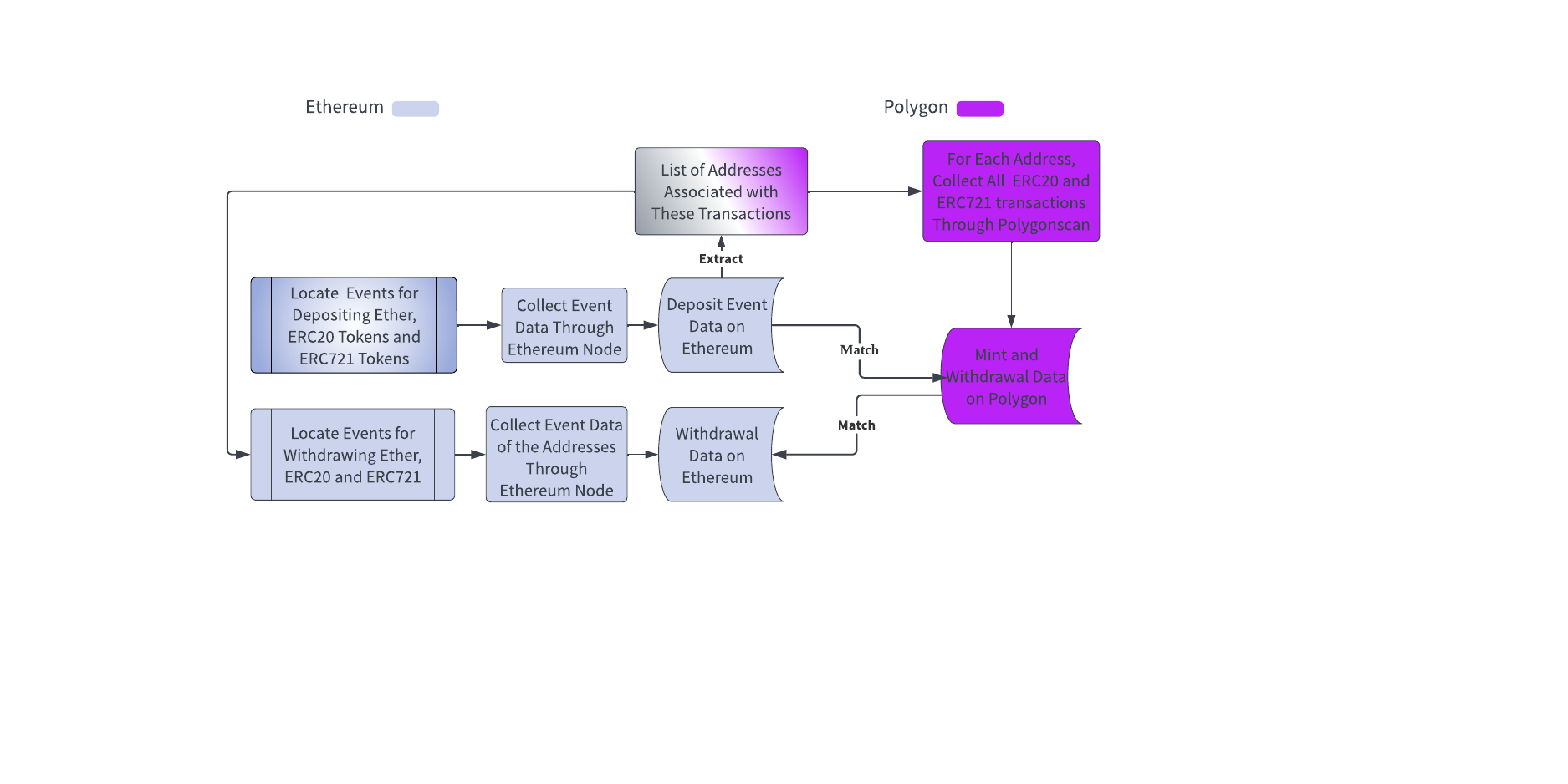}}
\caption{Pipeline for Polygon PoS Bridge data collection.}
\label{posdatacollection}
\end{figure}

In this section, we elaborate on how we collect data that will be used in our algorithm. Figure~\ref{posdatacollection} illustrates the data collection and matching pipeline for the Polygon PoS Bridge. Deposits from Ethereum to Polygon start with the asset-locking events on Ethereum, as the Polygon PoS Bridge utilizes the Lock-and-Mint mechanism. To start, we locate the events that lock the assets on Ethereum, namely \textit{LockedEther()}, \textit{LockedERC20()} and \textit{LockedERC721Batch()}.  These events correspond to depositing native Ether, ERC-20 tokens, and ERC-721 tokens, respectively, and each is associated with a smart contract on Ethereum. Next, we fetch the event logs using Ethereum's Erigon node and extract the desired data fields. Once we have all the data from Ethereum, we extract the list of addresses that have initiated deposit events on Ethereum. Since the Polygon POS Bridge uses  StateSync\footnote{https://wiki.polygon.technology/docs/pos/design/bridge/state-sync/state-sync/} to bridge data from Ethereum to Polygon, we cannot fetch Polygon event data because \textit{StateSync} is handled by Heimdall validators instead of contracts. Therefore, for each of these addresses, we query its full list of transactions of the corresponding token type on Polygon using Polygonscan's API\footnote{https://polygonscan.com/}. Then, for each event on the Ethereum side, we match it with its list of transactions on Polygon to reveal the cross-chain transactions. 

In addition to collecting the asset minting transactions on Polygon, this address list will also be used to collect withdrawal data on Ethereum. We only collect the withdrawal data of addresses that have initiated deposits before. Withdrawal data are more complex. Unlike deposits, all withdrawals are not associated with dedicated event types, but instead rely on standard \texttt{Transfer()} events, making them difficult to distinguish from regular token transfers. To address this, we apply a filtering strategy: we collect all transactions from the addresses that have deposit transactions before and retain only those with method ID \texttt{0x3805550f} in the \textit{input} field, which indicates a withdrawal from Polygon. 

Finally, the data we collected span from Aug 28\textsuperscript{th}, 2020 to Aug 28\textsuperscript{th}, 2023. The deposit and withdrawal events for Ethers, ERC20 tokens and ERC721 tokens across the Polygon POS Bridge are collected independently. In total, we have 1,528,318 records of Ether deposits, 558,190 records of ERC20 token deposits, and 34,135 records of ERC721 token deposits across the Polygon POS Bridge. As for withdrawals, we have 299,859, 293,688, and 5,925 records for Ethers, ERC20 tokens and ERC721 tokens, respectively. 

Notably, the number of total withdrawals data includes two scenarios:  (1) users withdrawing assets they had previously deposited; and (2) users initiating withdrawals without prior deposits. For instance, this pair of cross-chain transactions\cite{withdraw_not_deposit_pol,withdraw_not_deposit_eth} that withdrew 0.048 ETH from \textit{Polygon (Matic): Ether Bridge} contract but never received WETH  from the \textit{Null} contract on Polygon before, therefore it belongs to the second scenario. Regarding the matching withdrawal transactions, we consider the first scenario as cross-chain token withdrawal. From the withdrawal data, we filtered out transactions from addresses that have received tokens before from Ethereum. In total, there were 225,762 withdrawals for ethers, 264,173  for ERC20 tokens,  and 5,650 for ERC721 tokens. An overview of the collected data can be found in Table \ref{table:dataoverview}. 

\begin{table*}[htbp]
\caption{An overview of the collected data on Ethereum.}
\centering
\begin{tabular}{l @{\hspace{0.7cm}} l @{\hspace{0.7cm}} l @{\hspace{0.7cm}} l}
\toprule
\textbf{\makecell{Token\\ Type}} & \textbf{\makecell{Number of \\ Deposits}} & \textbf{\makecell{Number of \\Total  Withdrawals}} & \textbf{\makecell{Number of Withdrawals\\   by depositors}} \\
\midrule
Ether   & \hspace{1cm}1,528,318 & \hspace{1cm}299,859 & \hspace{1cm}225,762 \\
ERC20   & \hspace{1cm}558,190   & \hspace{1cm}293,688 & \hspace{1cm}264,173 \\
ERC721  & \hspace{1cm}34,315    & \hspace{1cm}5,925    & \hspace{1cm}5,650 \\
\bottomrule
\end{tabular}
\label{table:dataoverview}
\end{table*}

\section{Results}\label{sec:results}

\subsection{Evaluation of the match algorithm}
As the matched dataset serves as the foundation of this study, it is essential to evaluate the effectiveness of our cross-chain matching algorithm before presenting subsequent findings. The overall matching results, segmented by token type and transaction direction, are presented in Table~\ref{matchingratetable}. Based on this data, we draw several important observations:

\subsubsection{High overall match rates despite imperfections} Table \ref{matchingratetable} shows that the match rate for deposits and withdrawals of three types of tokens are over 80\% except the ERC20 withdrawals. While the ideal scenario would be a 1-to-1 match for every cross-chain transaction, our algorithm cannot achieve perfect accuracy due to several constraints:
\begin{itemize}

    \item Data completeness limitations: This limitation relays on the data source of Polygon. We use Polygonscan's API to collect data, for addresses with an extremely high volume of transactions on Polygon, the explorer API's inherent constraints may result in incomplete data collection. The API has a maximum query limit, for addresses with hundreds of thousands of transactions within a short timeframe, we may not capture their complete transaction history. For example, the address\footnote{\url{https://polygonscan.com/address/0xfA1fD291D6b235D32EaF4117058C824714c302f7}} demonstrates such high-frequency trading behavior. While an ideal approach to mitigate this limitation is to collect data from the Polygon node, the current implementation may omit certain Polygon transactions, leading to unmatched records.
    
    \item Time tolerance constraints: Another significant cause of mismatching is the time difference between transactions on Ethereum and Polygon. As described in Section~\ref{secmatch_algo}, transactions are only matched if they occur within a specified time window. However, cross-chain transactions involving multiple intermediate steps may exceed this time tolerance. In extreme cases, transaction propagation can take several hours, days, or even months. We present one example of this scenario (one on Ethereum\footnote{\url{https://etherscan.io/tx/0x52c3cd68ec95d3cb51c8cbbcfb772aad45251576af00a7494dc836e761fa2030}}, and another on Polygon\footnote{ \url{https://polygonscan.com/tx/0x0f7c92f0a6c38eef19dbe266264f611b4f340a7cc8381fab95325131bc8d4295}}), this cross-chain transaction took 6 months to complete. This proposed algorithm cannot capture transactions that take a longer time than the time tolerance.
    \end{itemize}
\subsubsection{High deposit matching rate but low withdrawal matching rate} 
The match rate data reveals a significant disparity between deposit and withdrawal transactions. All three token types (Ether, ERC20, and ERC721) achieve deposit match rates exceeding 93\%, demonstrating the algorithm's effectiveness for Ethereum-to-Polygon transactions. However, withdrawal match rates are consistently lower, with notable variation across token types (92.78\% for ERC721, 81.74\% for Ether, and only 67.55\% for ERC20). 

This discrepancy stems from several factors. First, the bridges employ asymmetric mechanisms: deposits use a straightforward "Lock-and-Mint" process, while withdrawals require the more complex "Burn-and-Prove" mechanism that needs more steps and more time. Second, the time difference is substantial, deposits typically complete within 10-20 minutes, while withdrawals require several hours, days, or even months, often exceeding the algorithm's time tolerance parameter. Third, deposit events on Ethereum have distinct identifiers (LockedEther(), LockedERC20()), whereas withdrawal events on Polygon use standard Transfer() events, and are more difficult to distinguish from regular transactions. Finally, as with deposits, data limitations on the Polygon side further exacerbate the issue, especially for high-frequency addresses.

\subsubsection{High match rate for ERC721 tokens but low match rate for ERC20 tokens} 
Among the three asset types, ERC721 tokens consistently demonstrate the highest match rates for both deposits (99.65\%) and withdrawals (92.78\%), while ERC20 tokens exhibit the lowest withdrawal match rate (67.55\%), despite having a strong deposit match rate (93.04\%). This token-specific disparity can be explained by two reasons. First, ERC721 tokens use dedicated ExitedERC721() events that provide clear token identification, whereas ERC20 tokens rely on standard Transfer() events that are indistinguishable from regular token transfers, making them hard to trace. Second, NFTs (ERC721 tokens) possess unique identifiers (token IDs) that enhance matching precision, while fungible ERC20 tokens lack this distinguishing characteristic.

\begin{table*}[htbp]
\caption{Cross-chain transaction match rate based on token types.}
\centering
\begin{tabular}{l c c}
\toprule
\textbf{Token Type} & \textbf{Match Rate of Deposits} & \textbf{Match Rate of Withdrawals} \\
\midrule
Ether  & 1,451,413 / 1,528,318 = 94.97\% & 184,541 / 225,762 = 81.74\% \\
ERC20  & 519,347 / 558,190 = 93.04\%  & 178,460 / 264,173 = 67.55\% \\
ERC721 & 34,194 / 34,315 = 99.65\%   & 5,242 / 5,650 = 92.78\% \\
\bottomrule
\end{tabular}
\label{matchingratetable}
\end{table*}

\subsection{Cross-chain transaction analysis} 
We analyze deposit and withdrawal transactions of Ether, ERC-20 tokens, and NFTs using three metrics: time cost, transaction counts, and transaction volumes. Additionally, we take KONGZ VX NFT as a case study to study if NFT transfers become more frequent or not after NFTs are bridged to Polygon as Polygon has lower transaction fees and higher speed.

\subsubsection{Time cost of cross-chain transactions}
\begin{figure}[htbp]
\centerline{\includegraphics[scale=0.3]{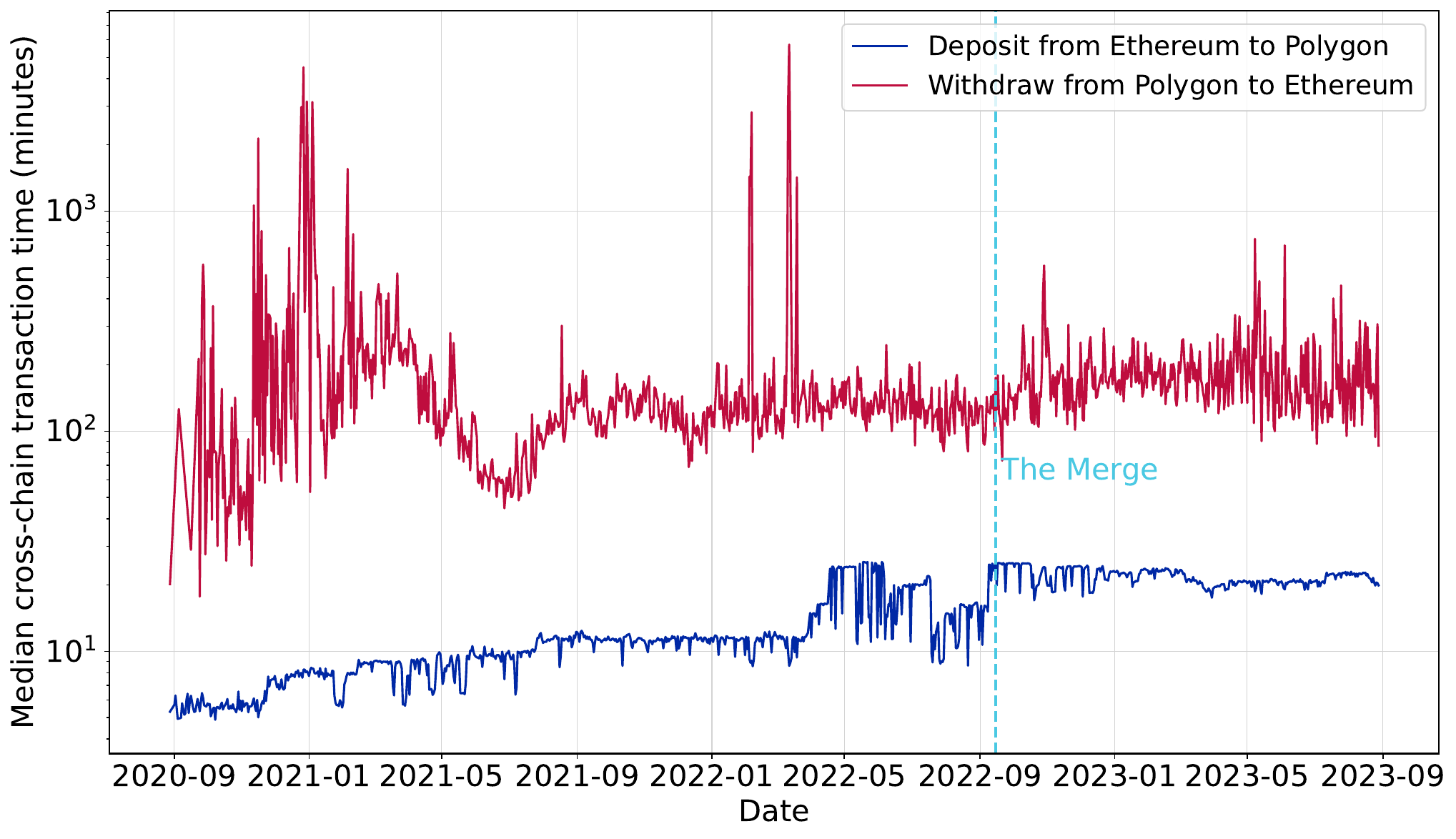}}
\caption{Time cost for Ether cross-chain transactions. }
\label{ethertime}
\end{figure}

\begin{figure}[t]
\centerline{\includegraphics[scale=0.3]{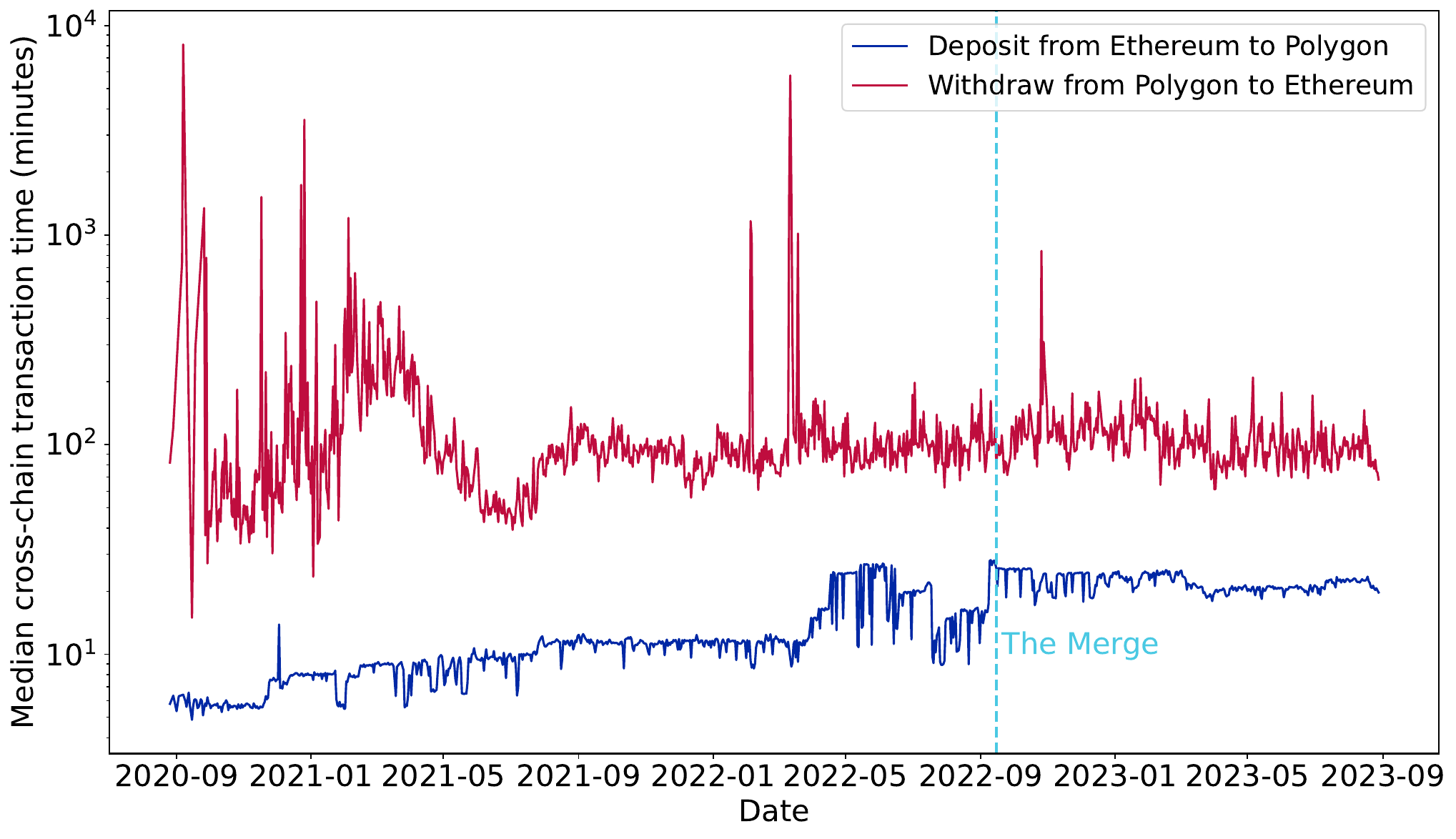}}
\caption{Time cost for ERC20 tokens cross-chain transactions.}
\label{erc20time}
\end{figure}

\begin{figure}[htbp]
\centerline{\includegraphics[scale=0.3]{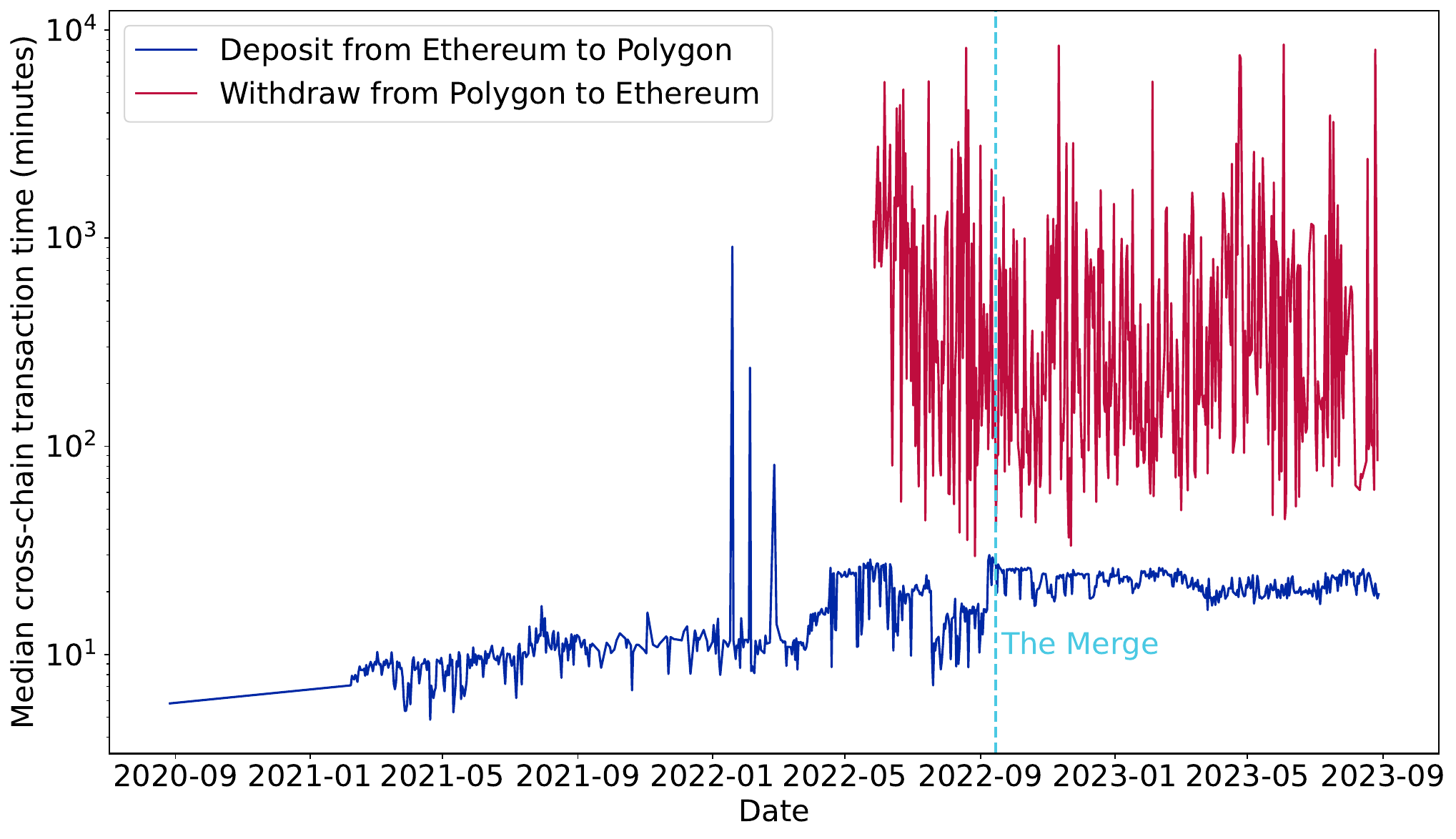}}
\caption{Time cost for NFT cross-chain transactions.}
\label{nfttime}
\end{figure}

Figures \ref{ethertime}, \ref{erc20time} and \ref{nfttime} illustrate the median time cost at a daily basis for Ether, ERC20 tokens and ERC721 tokens to travel between Ethereum and Polygon via the Polygon PoS Bridge. The time cost is defined as the difference in timestamps between the transactions on Ethereum and Polygon respectively. From the three plots, we have the following findings:
\begin{itemize}
    \item Before April 2022, it took about 10 minutes for all three types of assets to move from Ethereum to Polygon. However, after this point, the delay increased to approximately 20 minutes before decreasing again. Notably, after the Ethereum Merge, cross-chain transactions from Ethereum to Polygon increased and stabilized at around 20 minutes. This is due to Ethereum’s extended time to network finality after transitioning from PoW to PoS\cite{deposit_increaing_time}.
    \item Withdrawals from Polygon to Ethereum typically take significantly more time compared to deposits, regardless of the type of assets. the median value of Ether Withdrawals is 116 minutes, while the median value of ERC20 token withdrawals is around 95 minutes. However, withdrawals for ERC721 tokens have a larger time cost around 265 minutes to complete. Why withdrawals from Polygon to Ethereum take significantly longer than deposits is due to the mechanism and process differences. Deposits from Ethereum to Polygon utilize a Lock-and-Mint mechanism, where assets are locked in an Ethereum smart contract, and a corresponding amount is minted on Polygon. This process is relatively fast and users just need to issue one transaction on Ethereum and it is typically completed in approximate 10-20 minutes. In contrast, withdrawals from Polygon to Ethereum follow a Burn-and-Prove mechanism, which needs two manual actions by users, users need to first burn assets on Polygon, and later after the verification of validators on Polygon, users can claim assets on Ethereum by issuing another transaction\cite{withdraw_two_steps,Howtowit47:online}. Notably, this time cost of withdrawals highly depends on the users' side, if users delay initiating the second step, the withdrawal can take hours, days, or even months to complete.
    
    \item Security implications of operational asymmetry in withdrawals. Unlike deposits that need users to initiate one transaction, for withdrawals, users must first perform a burn operation on the Polygon side, followed by a claim operation on the Ethereum side. Since the process involves two separate user-initiated steps, users might be unaware of this or just forget to complete the second step. This could result in assets not being claimed promptly, or even being permanently forgotten. For example, one user initiated a withdrawal on Polygon involving over \$2 million worth of WETH, but failed to complete the claim step on Ethereum for six months. This resulted in the assets remaining idle and at risk of being forgotten. Eventually, a third party helped the user claim the funds\cite{LostinTr13:online}.
    \item Another security comes from the long-time cost of cross-chain transactions. Figures \ref{ethertime}, \ref{erc20time}, and \ref{nfttime} also show that some transactions took an extremely long time to across chains,  with one documented Ether withdrawal instance requiring 6 months to complete. These delays introduce multiple security vulnerabilities in the bridge governance architecture. First, validator centralization risk emerges as withdrawals depend on a limited set of validators reaching consensus\footnote{\url{https://staking.polygon.technology/validators}}, at the time of writing, there are 105 validators in total. Second, the extended asynchronous confirmation window creates opportunities for double-spending and replay attacks\cite{zhang2022xscope,zhang2024security} during the liminal state when assets are neither fully on the source nor destination chain. Third, the checkpoint submission frequency limitation results in assets remaining in an in-transit state, increasing liquidity freezing risk. Finally, network congestion produces amplification effects on cross-chain transactions. When the Ethereum network experiences congestion, the cost of checkpoint submissions by Polygon validators increases, particularly for withdrawals, potentially extending asset redemption periods. 
\end{itemize}

\subsubsection{The number of cross-chain transactions}
\begin{figure}[h]
\centerline{\includegraphics[scale=0.3]{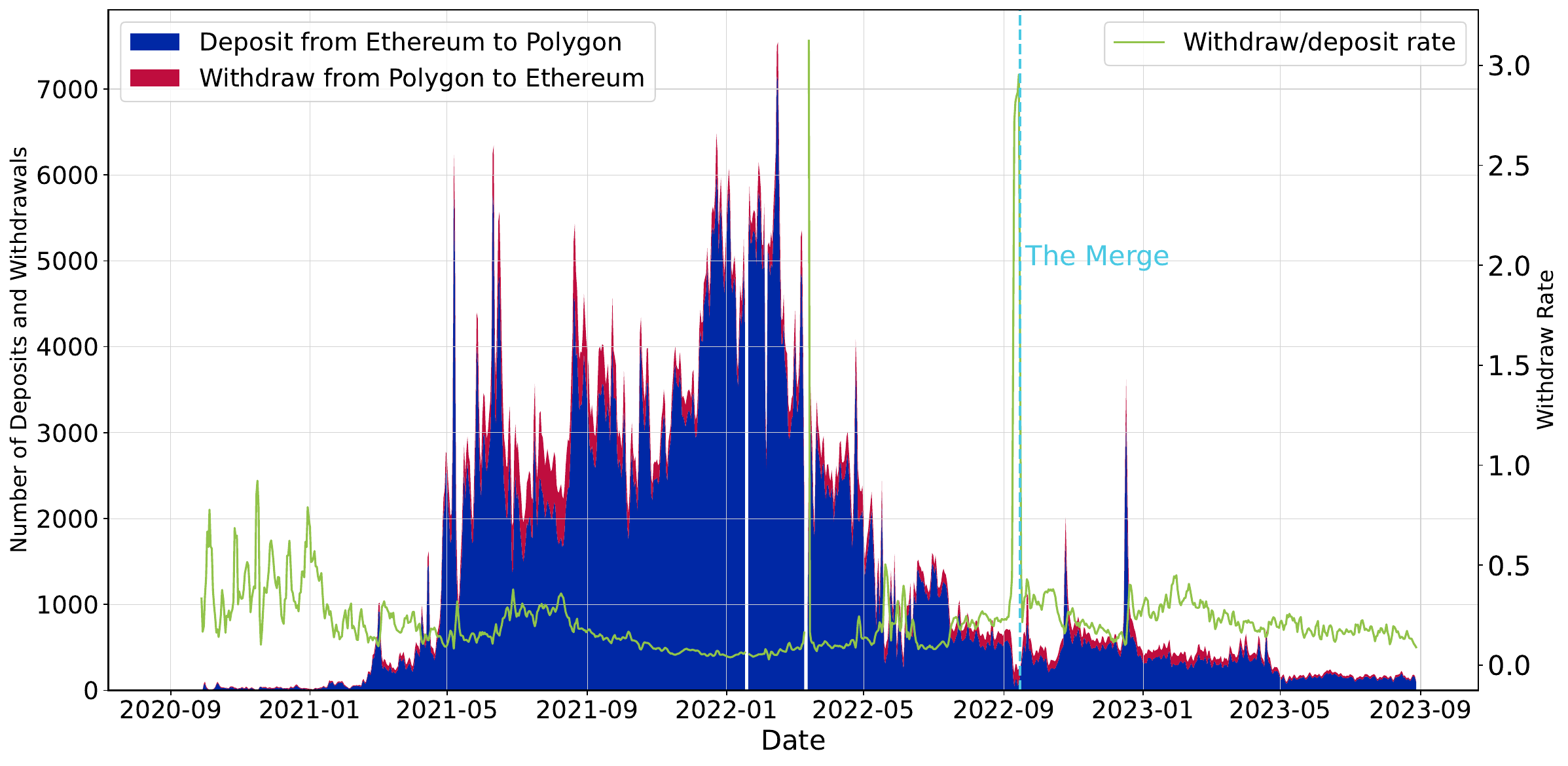}}
\caption{The number of deposits and withdrawals for Ether cross-chain transactions.}
\label{ethervolume}
\end{figure}

\begin{figure}[h]
\centerline{\includegraphics[scale=0.3]{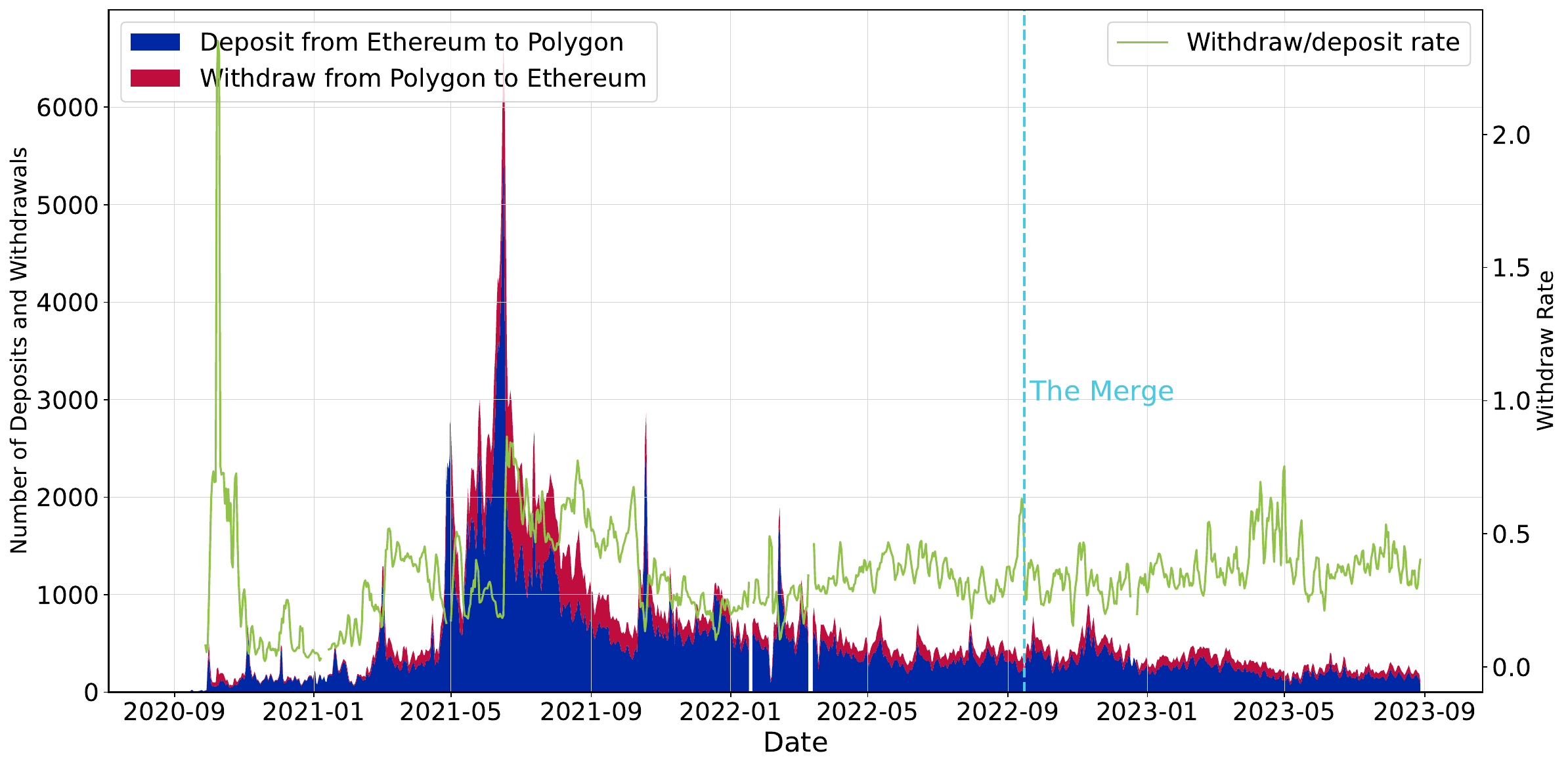}}
\caption{The number of deposits and withdrawals for ERC20 tokens cross-chain transactions. }
\label{erc20withdrawrate}
\end{figure}

\begin{figure}
\centerline{\includegraphics[scale=0.3]{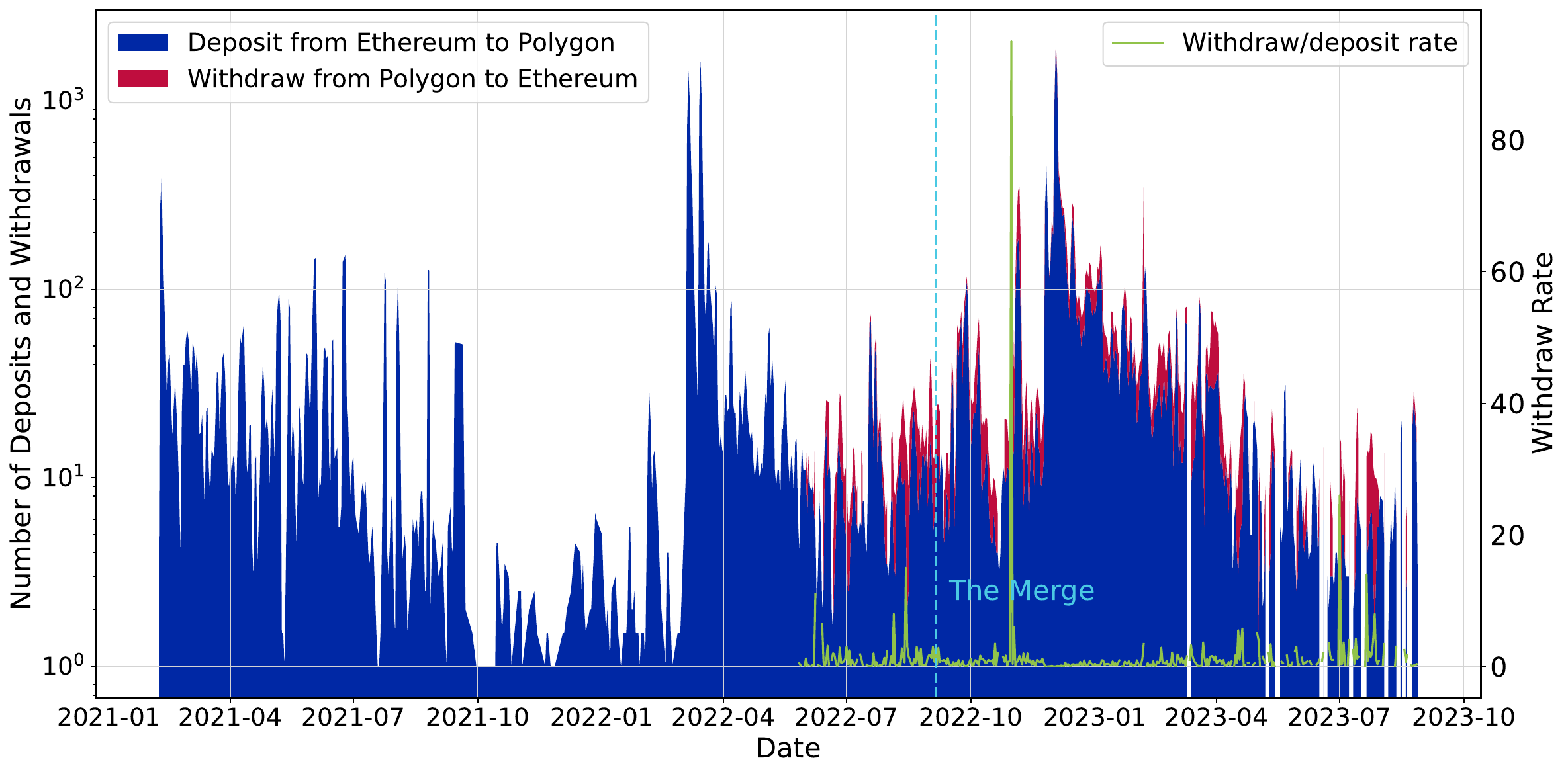}}
\caption{The number of deposits and withdrawals for NFTs cross-chain transactions.}
\label{nftnumberoftransactions}
\end{figure}

Figures \ref{ethervolume}, \ref{erc20withdrawrate}, and \ref{nftnumberoftransactions} illustrate the number of deposits, withdrawals, and withdrawal-to-deposit ratios for Ether, ERC20 tokens, and NFTs respectively over time. For Ether (Figure \ref{ethervolume}), transaction activity peaks around March 2022 with approximately 7,000 daily deposits, while ERC20 tokens (Figure \ref{erc20withdrawrate}) reach their maximum activity earlier, around June 2021 with approximately 6,000 daily transactions. In contrast, NFTs (Figure \ref{nftnumberoftransactions}) show a different pattern with their peak activity occurring much later around December 2022, though with significantly lower numbers. 

The withdrawal rates (green lines) reveal significant insights into cross-chain asset flow direction. For all three asset types, withdrawal rates typically remain below 50\%, indicating a dominant one-way movement from Ethereum to Polygon, with most users choosing to retain their assets on Polygon rather than transferring them back to Ethereum. This pattern aligns with Polygon's value proposition as a scaling solution with lower transaction costs and faster processing times. Notably, a notable spike in withdrawal-to-deposit ratios across all asset types on September 13, 2022, and this is directly attributed to Polygon's official announcement on September 12 regarding the Ethereum Merge\cite{TheMerge43:online}. In this announcement, Polygon explicitly stated that their bridging services would be temporarily paused during the Merge period. While smart contracts would remain active, the bridging interface would be unavailable until the Merge completed on September 15. The announcement advised users against bridging during this period. In response, users rushed to complete withdrawals from Polygon to Ethereum before the pause, while new deposits from Ethereum to Polygon sharply declined. The temporary pause of the Polygon bridge during Ethereum's consensus transition demonstrates how cross-chain bridges remain fundamentally dependent on their underlying blockchains' security mechanisms and operational states.


\begin{figure}[h]
\centerline{\includegraphics[scale=0.35]{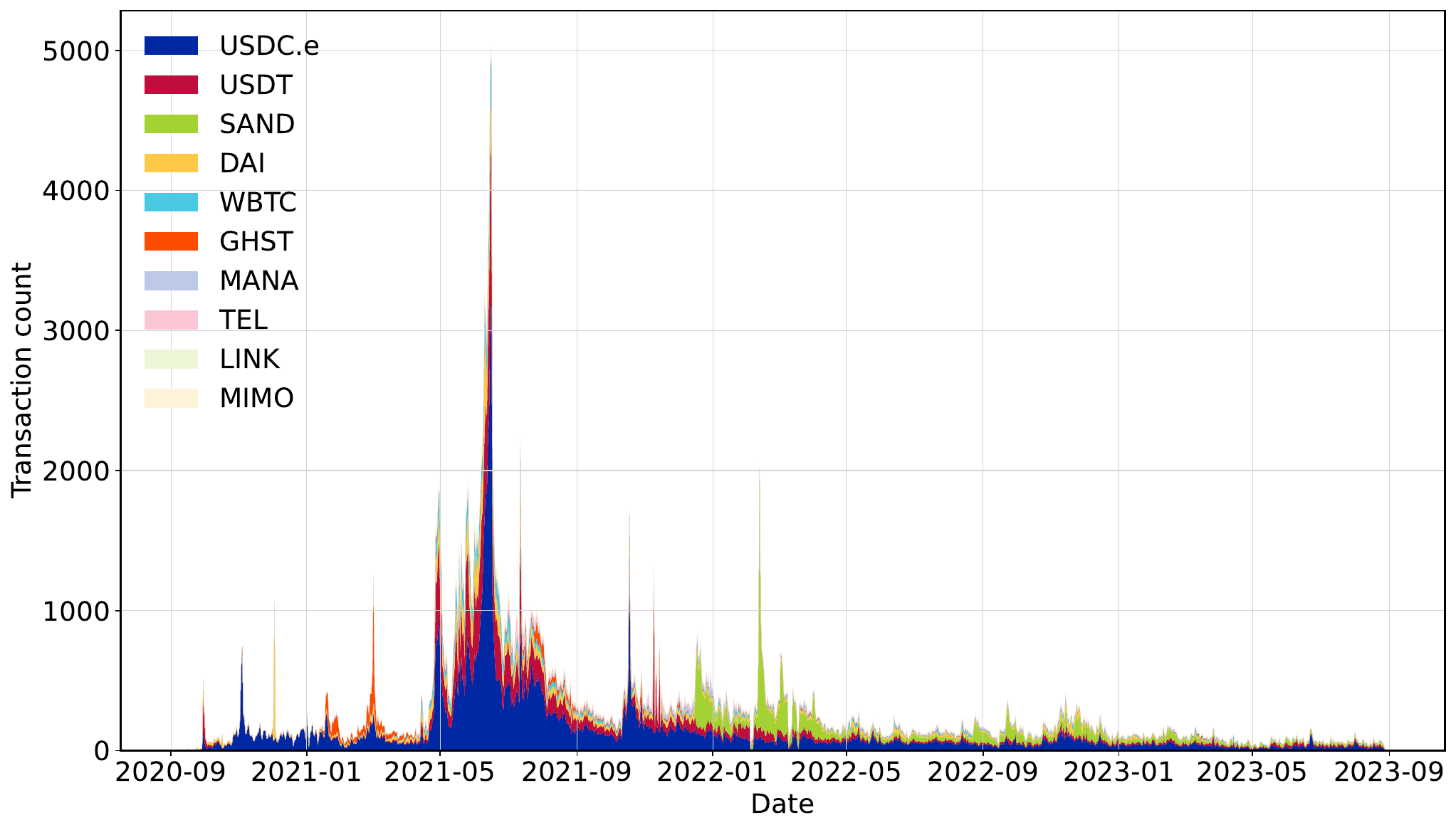}}
\caption{The number of cross-chain transactions for different ERC20 tokens.}
\label{erc20numberoftransactions}
\end{figure}

Figure \ref{erc20numberoftransactions} reveals distinctive patterns in ERC20 token cross-chain transactions between Ethereum and Polygon. The data clearly demonstrates that stablecoins, particularly USDC.e, USDT, and DAI dominate cross-chain activity in terms of transaction count. This prevalence highlights stablecoins' critical role as price-stable equivalents across blockchain ecosystems, serving as "digital dollars" that minimize volatility risk during cross-chain transfers.

\subsubsection{Transaction activities before and after cross-chain}

\begin{figure}[htbp]
\centerline{\includegraphics[scale=0.5]{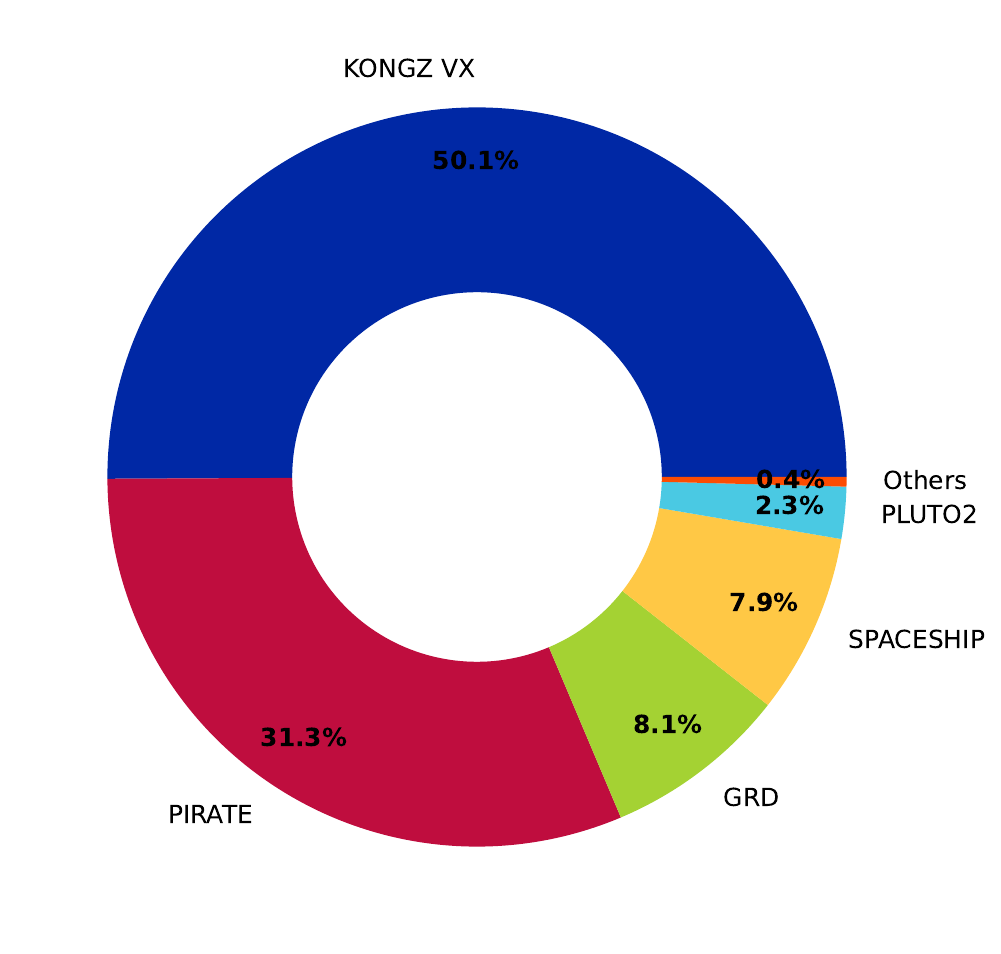}}
\caption{Main NFT collections bridged from Ethereum to Polygon.}
\label{nftcomposition}
\end{figure}

Users might be incentivized to bridge their assets from Ethereum to Polygon due to Polygon's lower transaction fees and faster confirmation times, which theoretically could lead to increased trading activity. To investigate this, we conduct a case study on KONGZ VX, the most frequently bridged NFT collection, which accounts for nearly 50\% of all ERC721 cross-chain transactions between Ethereum and Polygon (Figure \ref{nftcomposition}). We analyse the complete transaction network of KONGZ VX on both blockchains, covering both cross-chain and intra-chain transactions.

\begin{figure}[htbp]
\centerline{\includegraphics[scale=0.5]{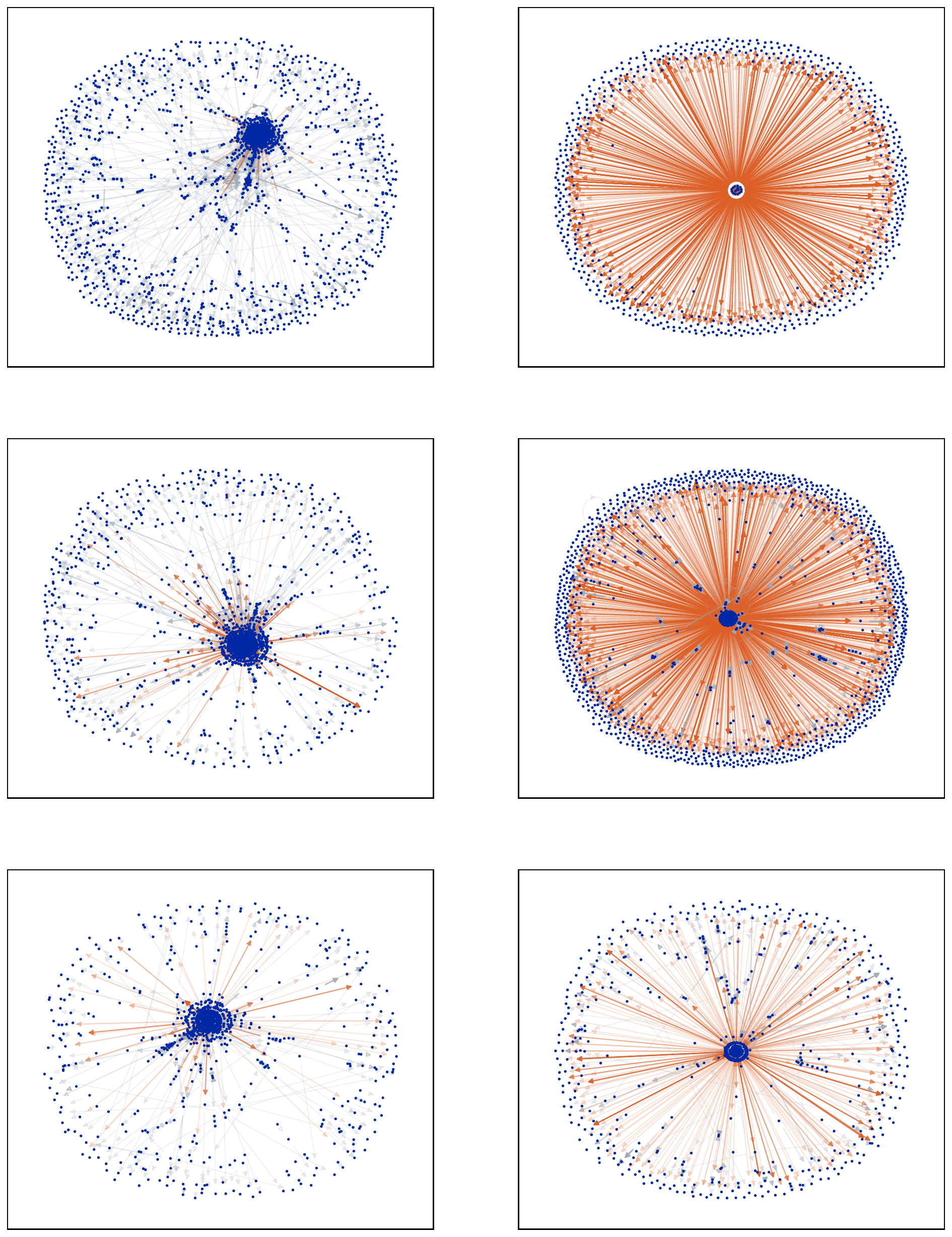}}
\caption{Network visualization of KONGZ VX. The left column represents the transaction network on Ethereum, and the right column represents the network on Polygon. From top to bottom, each row corresponds to March 2022, April 2022, and May 2022.}
\label{cyberkongzvis}
\end{figure}

Figure \ref{cyberkongzvis} presents a network visualization of KONGZ VX transactions across three key months with high activity (March-May 2022). Each row of networks corresponds to the Ethereum and Polygon transaction networks for a specific month. Each node in the networks corresponds to a unique address involved in transactions. An edge in the network represents a transaction from one node to another. Gray edges indicate transactions that do not cross the bridge, while orange edges denote cross-chain transactions. 
The visualization reveals a striking contrast between the two ecosystems. On Ethereum (left column), most of transactions are domestic transactions as reflected by the number of gray edges, while most of the transactions on Polygon are cross-chain transactions. This indicates that the transactions of KONGZ VX is still more active on Ethereum that on Polygon and NFT tokens are not being circulated after arriving on Polygon.

\begin{figure}[htbp]
\centerline{\includegraphics[scale=0.3]{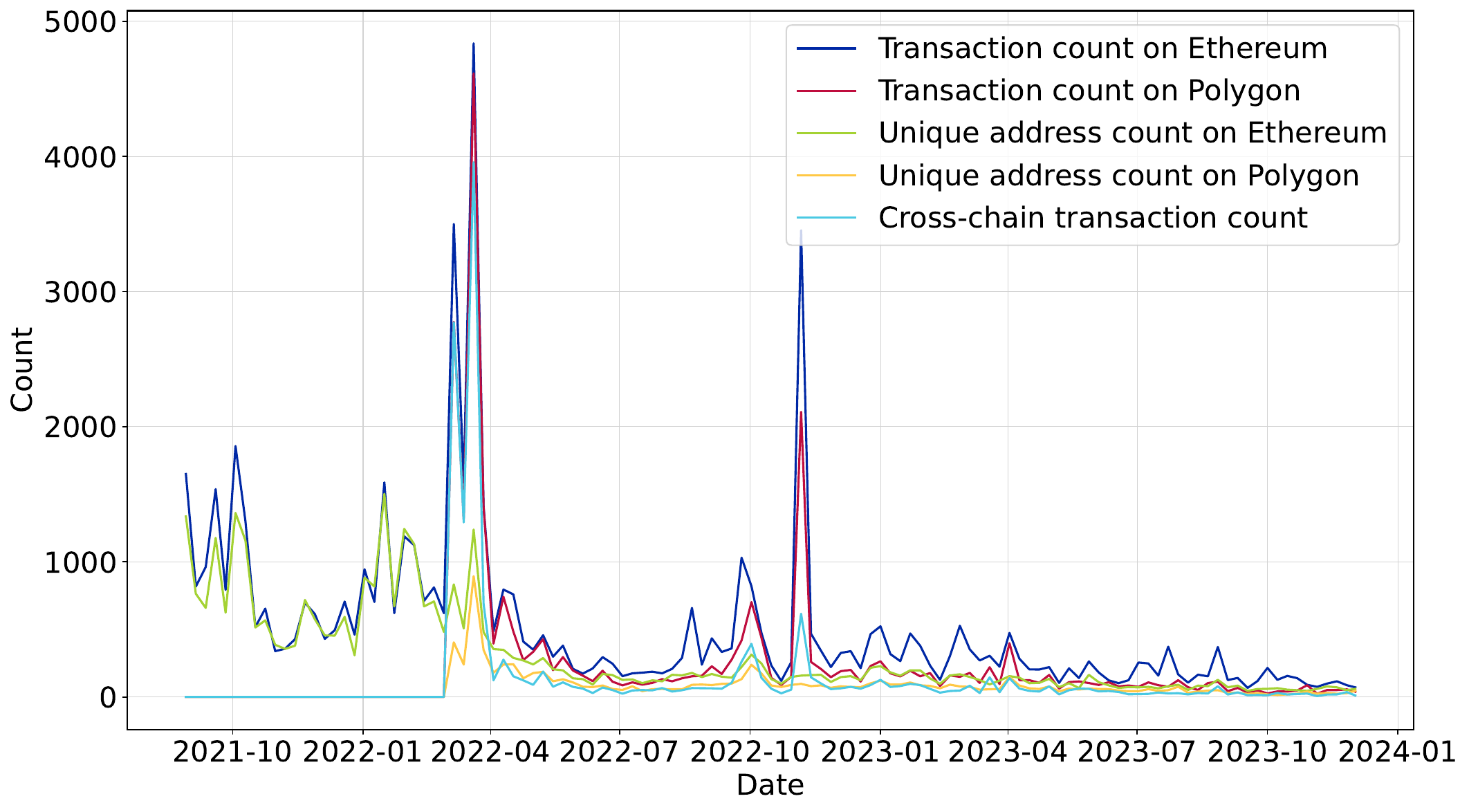}}
\caption{Active addresses and the number of transactions of  KONGZ VX NFT.} 
% Figure \ref{cyberkongzvis}.
\label{cyberkongzdata}
\end{figure}

\begin{figure}[htbp]
\centerline{\includegraphics[scale=0.3]{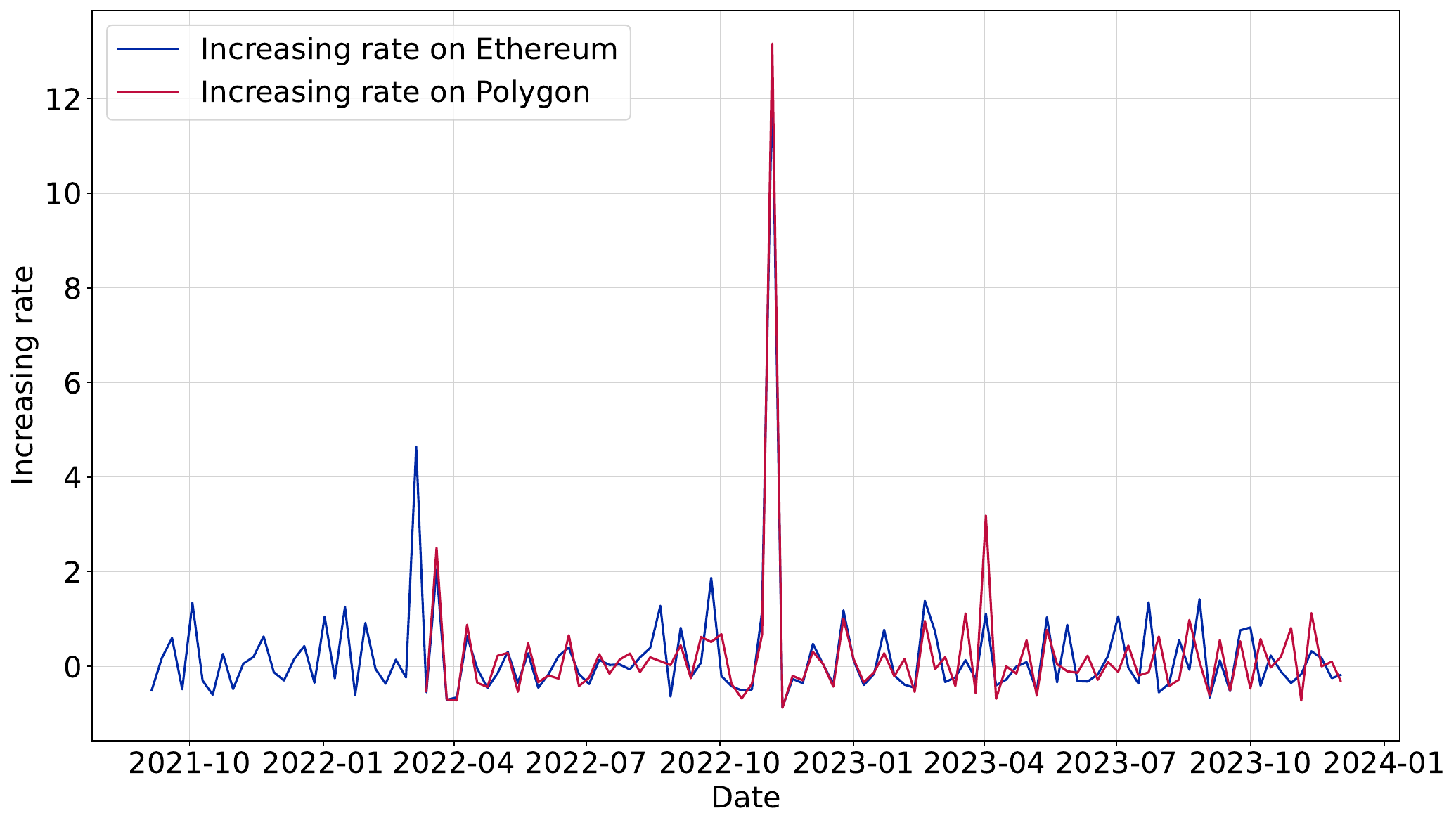}}
\caption{Increasing rate of the number of transactions of KONGZ VX NFT.}
\label{cyberkongzgrowthrate}
\end{figure}

\begin{figure}[htbp]
\centerline{\includegraphics[scale=0.3]{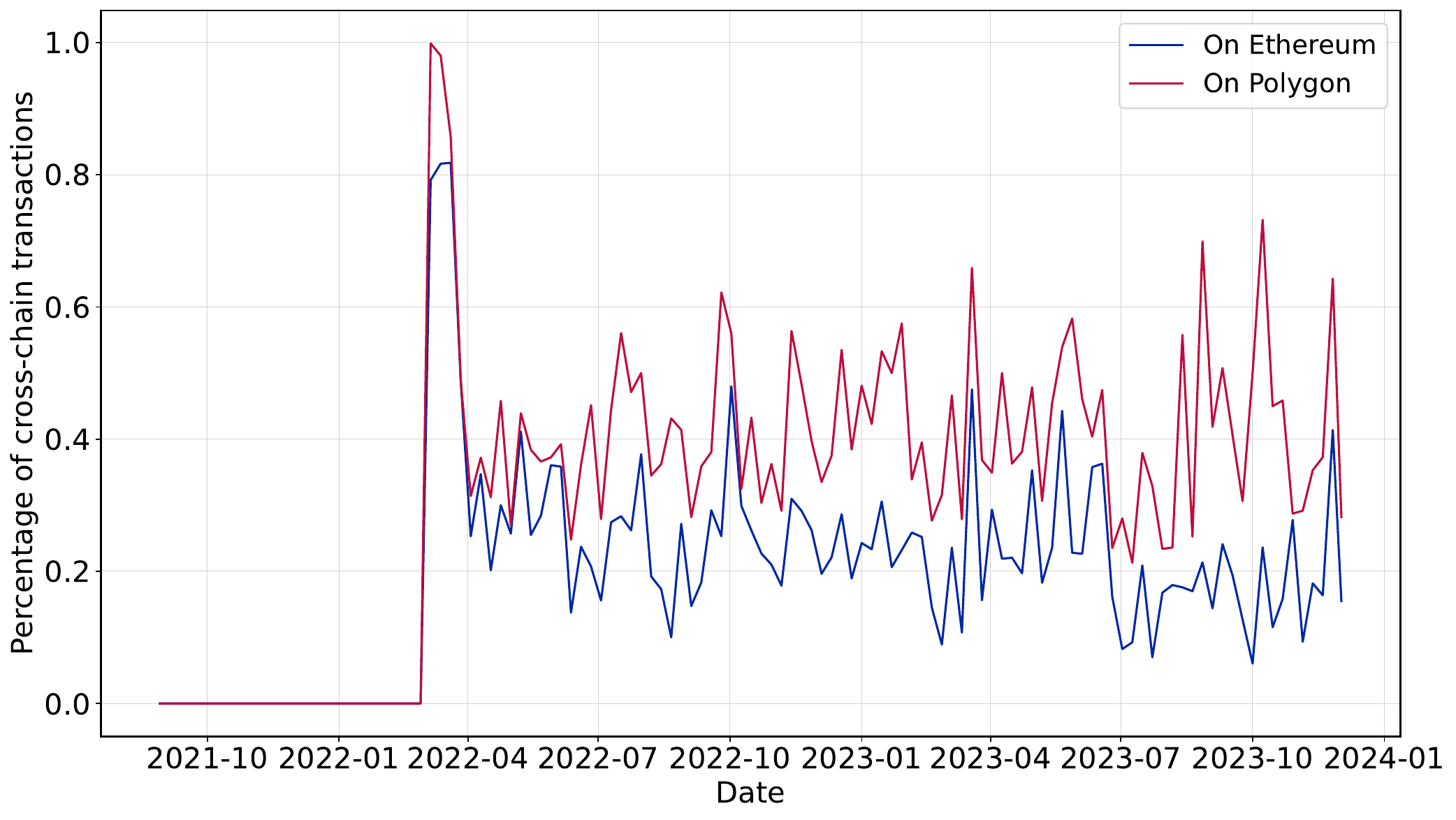}}
\caption{Percentage of cross-chain transactions of KONGZ VX NFT.}
\label{cyberkongzpercentage}
\end{figure}

Quantitative analysis in Figure \ref{cyberkongzdata} confirms this pattern, showing consistently a higher transaction count and more active addresses on Ethereum compared to Polygon. Despite Polygon's growth rate being slightly higher (Figure \ref{cyberkongzgrowthrate}), the absolute activity remains substantially lower. Figure \ref{cyberkongzpercentage} further reveals that the vast majority of Polygon transactions are cross-chain related rather than secondary trading. 

This pattern contradicts the expected behavior that lower transaction costs would stimulate more active transaction activities. Several factors may explain this phenomenon: (1) the Ethereum NFT market offers stronger liquidity and more established trading infrastructure; (2) NFTs bridged to Polygon enter a more fragmented ecosystem with fewer active traders; and (3) the added complexity and security concerns \cite{zhang2022xscope,zhang2024security} of cross-chain transactions may discourage active traders from operating on secondary chains.

These findings suggest that while Polygon provides cost advantages for NFT transactions, this benefit alone is insufficient to shift significant trading activity away from Ethereum's ecosystem. The persistence of trading concentration on Ethereum despite higher costs indicates that factors beyond transaction fees, including market liquidity, ecosystem maturity, and perceived security play crucial roles in determining where digital asset transaction activity concentrates.

\section{Discussion}\label{sec:dis}

\subsection{Possibility to extend this algorithm to Non-EVM based blockchains}
our methodology may be extended to non-EVM chains with some adaptations. While three of our matching criteria: value consistency, token identification matching, and temporal proximity have potential applicability across non-EVM blockchain, the core challenge lies in address consistency. Unlike EVM chains where identical addresses can be used across networks, non-EVM chains (such as Bitcoin with its Base58 format) utilize fundamentally different address generation systems that don't naturally correspond to Ethereum's hexadecimal addresses. Without this built-in address consistency mechanism, the search scope becomes prohibitively large, making accurate matching difficult. However, if the bridge on a non-EVM blockchain records the address mapping or explicitly includes the recipient address on the EVM-based chain, our method can still be applied in such cases. Additionally, structural differences between account-based models (Ethereum) and UTXO-based systems (Bitcoin) would necessitate fundamental algorithmic redesign. Although our current approach traces one-to-one value transfers, UTXO systems would require analyzing complex input-output relationships that could split or combine values across multiple addresses.

\subsection{Governance model and security implications}
The Polygon PoS Bridge operates through a semi-centralized governance model with approximately 105 validators (as of March 2025) responsible for transaction validation and checkpoint submission. This relatively small validator set creates potential bottlenecks affecting transaction finality and security. Governance decisions have directly impacted cross-chain behavior, as evidenced by the bridge suspension during the Ethereum Merge that triggered significant behavioral shifts in user activity (Figure \ref{ethervolume}). The extended confirmation windows observed in withdrawals (95-265 minutes median, extremes reaching 6 months) expose users to governance-related risks including potential transaction censorship, but this needs to be investigated further. Unlike the relatively automated deposit process, withdrawals introduce multiple governance-dependent steps where failures or delays may arise. This interplay between governance and security is particularly significant, as users often assume equivalent security properties across both blockchains involved in a cross-chain transaction. In practice, however, the security of such transactions depends not only on the underlying protocols of the source and destination chains, but also on the technical architecture and governance structure of the bridge itself.

\section{Conclusion}\label{sec:con}

In this work, we developed a novel heuristic algorithm for tracking cross-chain transactions between EVM-compatible blockchains, achieving high matching rates for deposits while revealing lower effectiveness for withdrawals. Our analysis revealed significant temporal asymmetries between deposits and withdrawals, with extreme delays exposing security vulnerabilities including validator centralization risk and double-spending attacks during extended confirmation windows. Additionally, we observed the Ethereum Merge significantly increase the deposit completion times, demonstrating how consensus changes directly impact cross-chain operations. Our analysis of transaction patterns revealed predominantly one-way asset flows from Ethereum to Polygon (withdrawal-deposit rates below 50\% at most of the time), while stablecoins dominated ERC20 cross-chain activity. In contrast to theoretical expectations, our case study on the KONGZ VX NFT collection demonstrated that, despite Polygon’s lower transaction fees, user activity remained concentrated on Ethereum. This suggests that factors such as market liquidity, ecosystem maturity, and perceived security may outweigh cost advantages when it comes to digital asset trading behaviour.

While our study offers valuable insights into cross-chain transactions between Ethereum and Polygon, several key limitations must be acknowledged. First, our matching algorithm relies on address consistency across EVM-compatible blockchains, restricting its applicability to non-EVM chains with different address formats or transaction structures. Second, although our algorithm has a high matching rate for deposit transactions, the matching rate for withdrawal transactions is still relatively low. Third, the transaction data on the Polygon side relies on explorer APIs may lead to incomplete transaction histories, particularly for addresses with high transaction volumes due to API query limits. lastly, our analysis focuses specifically on the Polygon PoS Bridge, which may exhibit unique characteristics not found in other cross-chain bridges, limiting the generalizability of our findings to different bridge implementations or blockchain pairs. For future research, firstly, our methodology can be extended to analyze cross-chain transactions between other EVM-compatible blockchains, such as Ethereum and Arbitrum or Optimism, offering broader insights into cross-chain ecosystems. Additionally, our matched transaction dataset enables a deeper exploration of cross-chain arbitrage and MEV activities\cite{yan2024optimizing,zhang2024improved,mcmenamin2023sok,gogol2024cross,gogol2024layer}, including arbitrage patterns and complex trading behaviors.

\bibliography{main}

\end{document}